\documentclass[aps,prd,          groupedaddress,showpacs]{revtex4}

\usepackage{epsfig}

\begin{document}


 \def\lp{\left. }
 \def\rp{\right. }
 \def\lr{\left( }
 \def\rr{\right) }
 \def\le{\left[ }
 \def\re{\right] }
 \def\lg{\left\{ }
 \def\rg{\right\} }
 \def\lb{\left| }
 \def\rb{\right| }

 \def\li{\mbox{Li}_2}

 \newcommand{\ms}{m_{\tilde{q}}}
 \newcommand{\mc}{m_{\tilde{\chi}}}
 \newcommand{\mg}{m_{\tilde{g}}}
 \newcommand{\nonu}{\nonumber \\}

\def\bea{\begin{eqnarray}}
\def\eea{\end{eqnarray}}
\def\ppbar{{\rm p} \bar{{\rm p}}}
\def\pp{ {\rm p} {\rm p} }
\def\ifb{ {\rm fb}^{-1} }
\def\del{\partial }
\def\ra{\rightarrow}
\def\Ra{\Rightarrow}
\def\dis{\displaystyle}
\def\f{\frac}

\def\chiplus{{\tilde{\chi}}^{+}}
\def\chiminus{{\tilde{\chi}}^{-}}
\def\chizero{{\tilde{\chi}}^{0}}
\def\chargino{{\tilde{\chi}}^{\pm}}
\def\neutralino{{\tilde{\chi}}^{0}}
\def\gaugino{{\tilde{\chi}}}
\def\gluino{\tilde{g}}
\def\bino{\tilde{B}}
\def\w3ino{\tilde{W_3}}
\def\h1ino{\tilde{H_1}}
\def\hino2{\tilde{H_2}}
\def\wpino{\tilde{W}^{+}}
\def\hpino{\tilde{H}^{+}}
\def\bbar{\bar{b}}
\def\tbar{\bar{t}}

\def\ifb{${\rm fb}^{-1}$}
\def\alphas{\alpha_{S}}
\def\alphash{\hat{\alpha}_{S}}
\def\MS{$\overline{\rm MS}$\,\,}

\def\u6d{u_{6 \Delta}}
\def\ud7{u_{7 \Delta}}
\def\s4d{s_{4 \Delta}}
\def\sd3{s_{3 \Delta}}
\def\delu{\Delta_{u}}
\def\delt{\Delta_{t}}

\newcommand{\lae}{\stackrel{<}{\sim}}
\newcommand{\gae}{\stackrel{>}{\sim}}
\newcommand{\msqu}[1]{m_{\tilde{q}_{#1}}}
\newcommand{\ih}[2]{\,\hat{I}\left( \frac{#1}{#2} \right)}

\preprint{hep-ph/0212306}

\title{Erratum: Next-to-leading order supersymmetric QCD predictions for
 associated production of gauginos and gluinos [Phys.\ Rev.\ D 62, 095014
 (2000)]}

\author{Edmond L.\ Berger, Michael Klasen, and Tim M.\ P.\ Tait}

\date{\today}


\maketitle


Within the curly brackets of Eq.\ (C1), a term $-\pi^2/4$ from the expansion of
the relative factor $\Gamma(1-\epsilon)/\Gamma(1-2\epsilon)$ between the
virtual (Eq.\ (21)) and the soft corrections (Eq.\ (28)) should be included.
Equation (C1) then reads
\renewcommand{\theequation}{C1}
\bea
   \frac{d^2\hat{\sigma}^S}{dt_2 \, du_2} &=& 
   \frac{d^2 {\hat{\sigma}}^B}{dt_2 \, du_2} 
   \left( \frac{C_F \, \alphas}{\pi} \right) \left\{
     {\rm Li}_2 \left( \frac{ u_2 \, t_2 - s \, m_2^2}{(s+t_2)(s+u_2)} \right) 
   + \frac{1}{2} \log^2 \left( \frac{ \mu^2}{m_1^2 \delta^2} \right)
   + \log \left(\frac{(s+t_2)(s+u_2)}{s \, m_1^2} \right)
     \log \left( \frac{ \mu^2}{m_1^2 \delta^2} \right) \right. 
   \nonumber \\[0.3cm]
   & & \left.
   + \frac{1}{2}\log^2 \left( \frac{(s+t_2)(s+u_2)}{s \, m_1^2} \right)
   - {\pi^2\over4} \right\}.
\eea
Correct inclusion of the missing term changes the NLO cross sections and
their renormalization and factorization scale dependences. As examples,  
Figs.\ 16 and 17 should be replaced by Fig.\ 1 below. The corrected
%
\begin{figure}
 \centering
 \includegraphics[width=0.49\columnwidth]{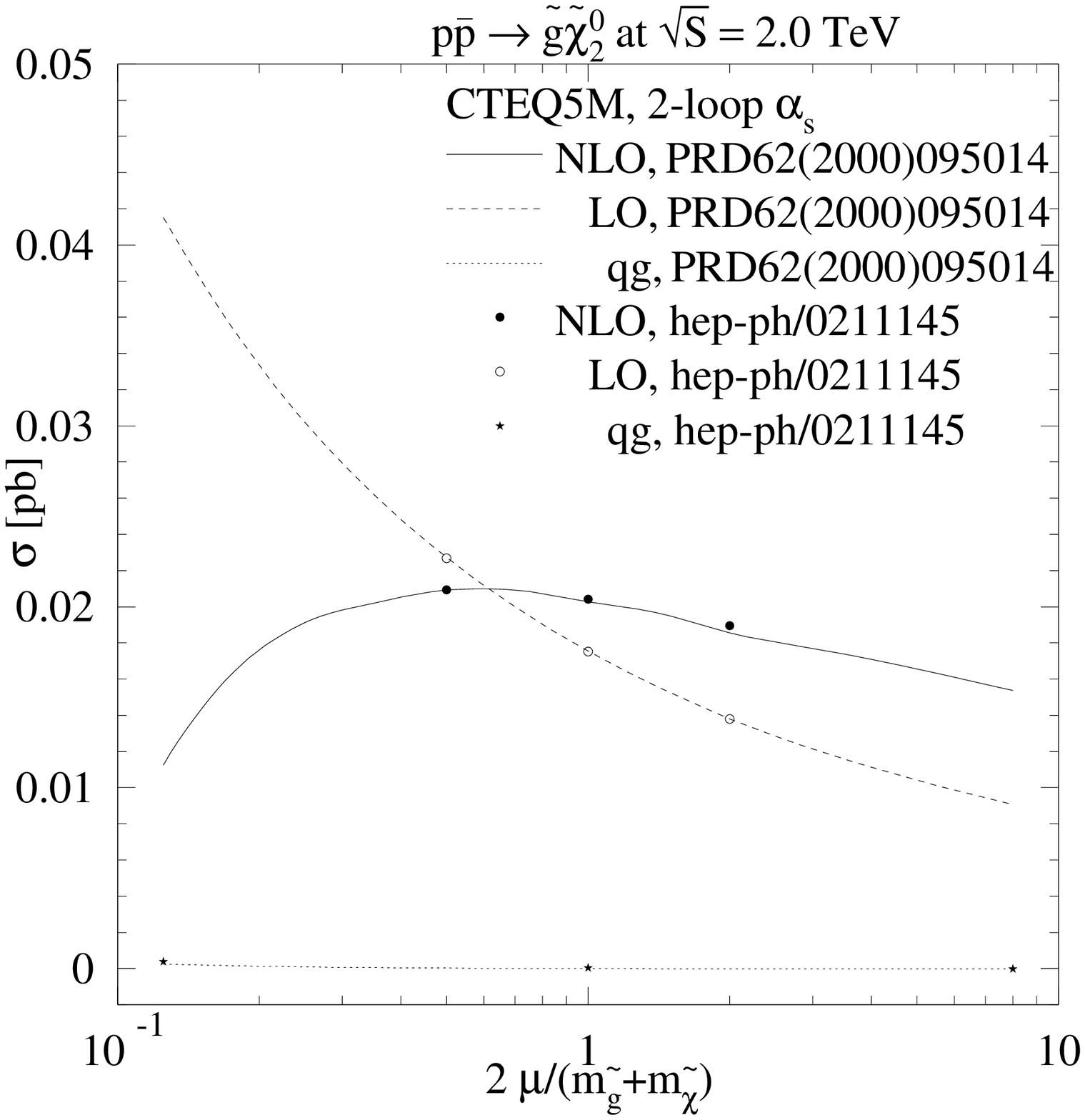}
 \includegraphics[width=0.49\columnwidth]{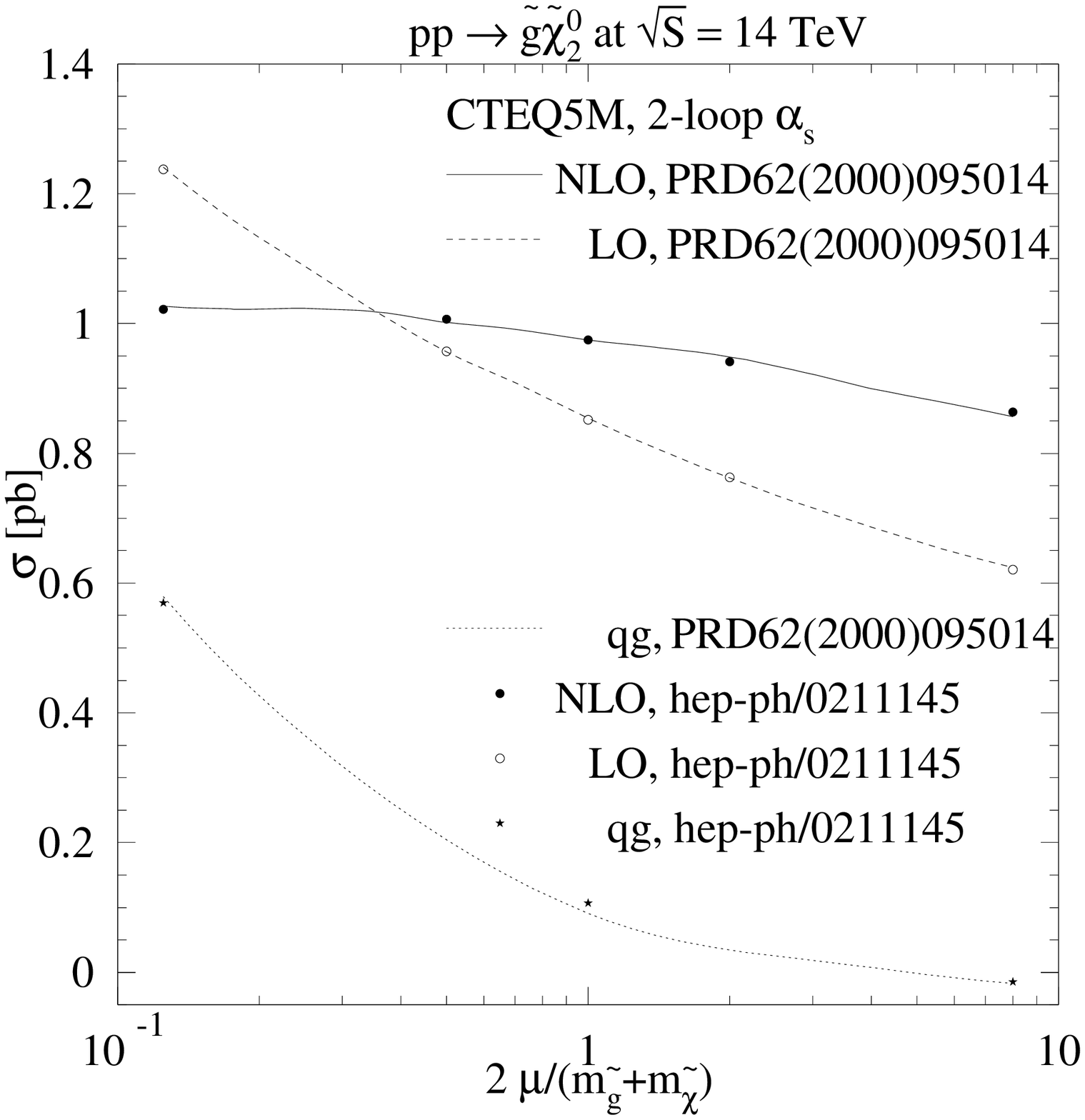}
 \caption{\label{fig:1}Dependence of the predicted NLO, LO, and $qg$ initiated
  total cross sections at the Tevatron (left) and LHC (right) on the
  renormalization and factorization scale. We show the case of $\gluino
  \neutralino_2$ associated production in the SUGRA model, with $m_{\gluino} =
  410 $ GeV and $m_{\neutralino_2} = 104$ GeV. This figure replaces Figs.\ 16
  and 17 in Ref.\ \cite{Berger:2000iu} and agrees with the results in Ref.\
  \cite{Spira:2002rd}.}
\end{figure}
%
results agree with those in Ref.\ \cite{Spira:2002rd}, if CTEQ5M parton 
densities \cite{Lai:1999wy} are used along with the two-loop expression for 
$\alpha_s$ and $\Lambda^{(5)}=226$ MeV. 

Our implementation of the intermediate on-shell squark subtraction in 
Eq.\ (41) differs from that in Ref.~\cite{Spira:2002rd}. While in Ref.[1] 
resonance contributions are subtracted strictly on-shell, small off-shell 
contributions from phase space are included in our subtraction procedure. 
The differences are insignificant quantitatively in this case.

Specific predictions for the Tevatron and LHC will be provided on request
\cite{request}. For example, we show in Fig.\ 2 the predicted total cross
sections at the current Tevatron center-of-mass energy of
%
\begin{figure}
 \centering
 \includegraphics[width=0.49\columnwidth]{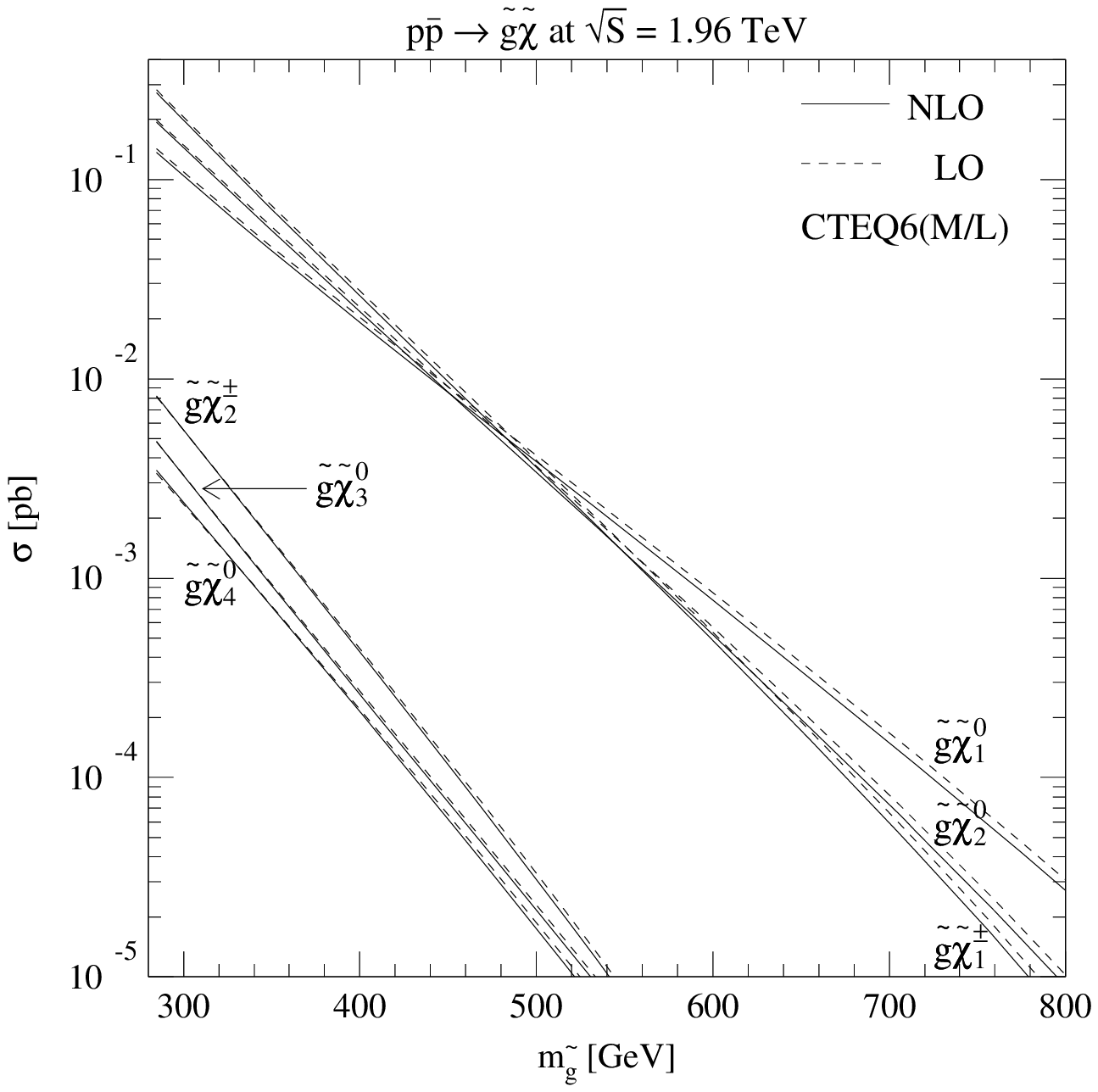}
 \includegraphics[width=0.49\columnwidth]{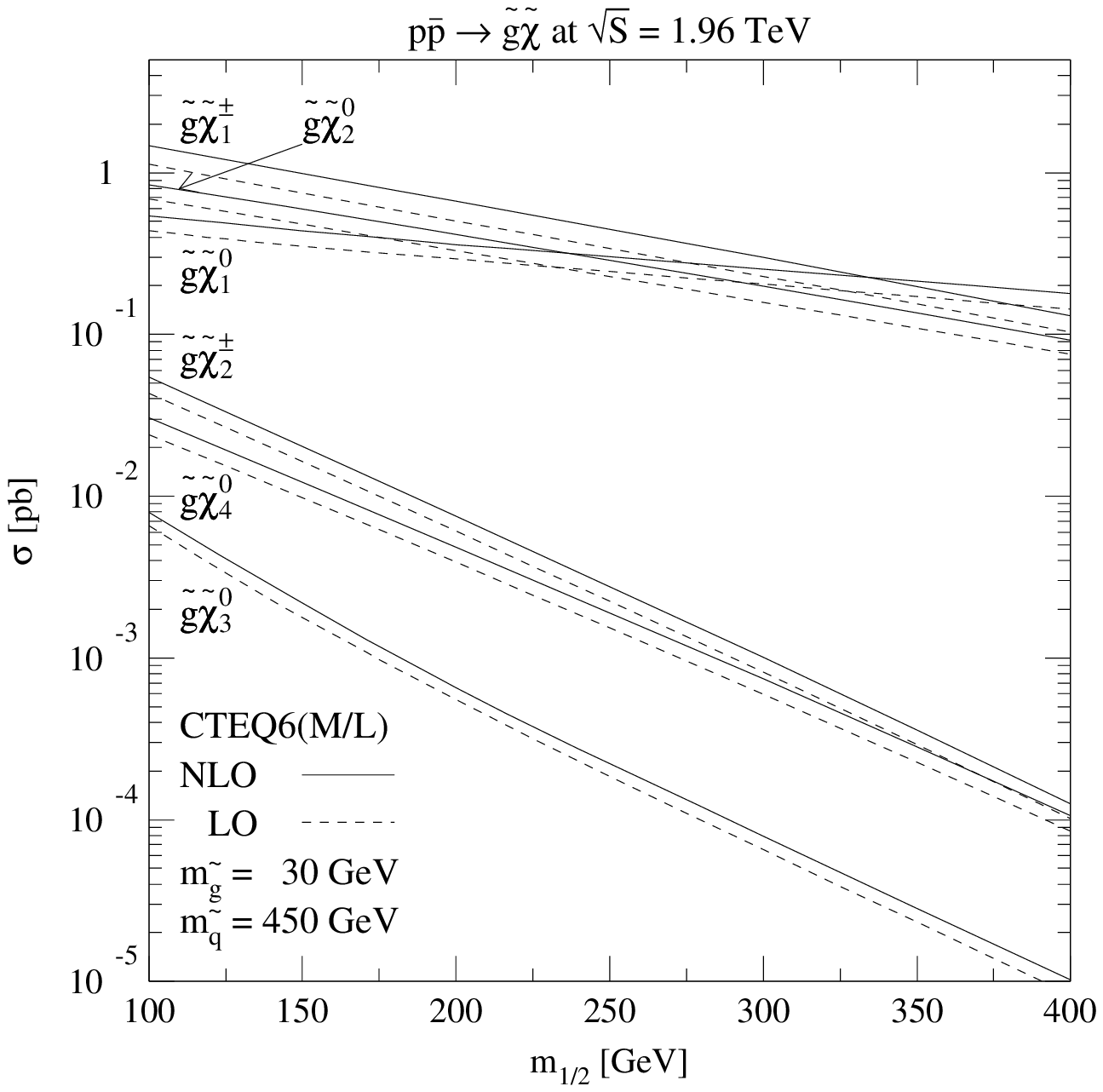}
 \caption{\label{fig:2}Predicted total cross sections at the Tevatron with 
  total center-of-mass energy $\sqrt{S}=1.96$ TeV and CTEQ6 parton densities 
  for all six $\gluino\gaugino$ channels in the SUGRA model as functions of
  the mass of the gluino (left) and for a gluino with mass 30 GeV as functions
  of $m_{1/2}$ (right).}
\end{figure}
%
$\sqrt{S}=1.96$ TeV for all six $\gluino \gaugino$ channels in the SUGRA model
as functions of the mass of the gluino (left) and for a gluino with mass 30 GeV
as functions of $m_{1/2}$ (right). In Fig.\ 2, we use CTEQ6M (NLO) and CTEQ6L
(LO) parton densities \cite{Pumplin:2002vw} and the two-loop expression for
$\alpha_s$ with $\Lambda^{(5)}=226$ MeV (LO {\em and} NLO).

Finally, we take the opportunity to correct two typographical errors in the
published version of Ref.\ \cite{Berger:2000iu}. In Eq.\ (C2), the arguments
of the squared and the last logarithm were interchanged. The correct form is
\renewcommand{\theequation}{C2}
\bea
   \frac{d^2\hat{\sigma}^S}{dt_2 \, du_2} &=& 
   -\frac{d^2 {\hat{\sigma}}^B}{dt_2 \, du_2} 
   \left( \frac{N_C \, \alphas}{2 \pi} \right) \left\{
     -2 + 
     {\rm Li}_2 \left( \frac{ u_2 \, t_2 - s \, m_2^2}{(s+t_2)(s+u_2)} \right) 
   + \frac{1}{2} \log^2 \left( \frac{(s+t_2)(s+u_2)}{s \, m_1^2} \right)
  \right. \nonumber \\[0.3cm] & & \left.
      + \left[ \log \left(\frac{(s+t_2)(s+u_2)}{s \, m_1^2} \right)
     - 1 \right]
     \log \left( \frac{ \mu^2}{m_1^2 \delta^2} \right) \right\}.
\eea
In Eq.\ (D2), $+$-signs were omitted between the two large brackets in the
third line and the first two fractions in the fourth line. The correct form is
\renewcommand{\theequation}{D2}
\bea
  \frac{d^3 \hat{\sigma}_1^{g}}{ds_4 \, dt_2 \, du_2} &=&
  \frac{ C_F \, \alphas \, \alphash \, \delta \, (s + t_2 + u_1 - s_4)}
  {36 \, \pi \, s^2}
  \log \left( \frac{\mu^2 (s_4 + m_1^2)}{s_4^2} \right) \\[0.3cm]
  & & \times \, \left\{
  \left( \frac{s_4^2 - 2 \, s_4 (s+u_2) + 2 (s+u_2)^2}{s_4 (s+u_2)} \right) 
  \, \left( \frac{X_t \, t_2}{(t - \msqu{t}^2)^2}
  + \frac{2 \, X_{t u} \, s \, m_1 \, m_2}
  {(t- \msqu{t}^2) [(\delu-s-t_2)(s+u_2) + s \, s_4]} 
  \right. \right. \nonumber \\[0.3cm]
  & & \left. \left.
  + \frac{X_u \, u_2 \, [s_4 \, u_2 - u_1 \, (s+u_2)]}
  {[(\delu-s-t_2)(s+u_2) + s \, s_4]^2} \right) \right.
+
  \left( \frac{s_4^2 - 2 \, s_4 (s+t_2) + 2 (s+t_2)^2}{s_4 (s+t_2)} \right) 
 \nonumber \\[0.3cm]
   & &  \left.
  \times \,\left( \frac{X_t \, t_2 \, [s_4 \, t_2 - t_1 \, (s+t_2)]}
  {[(\delt-s-u_2)(s+t_2) + s \, s_4]^2} +
  \frac{2 \, X_{t u} \, s \, m_1 \, m_2}
  {(u- \msqu{u}^2) [(\delt-s-u_2)(s+t_2) + s \, s_4]}
  + \frac{X_u \, u_2}{ (u - \msqu{u}^2)^2 }
  \right) \right\} . \nonumber
\eea



\end{document}